\documentclass[sigconf]{acmart}
\usepackage{graphicx}
\usepackage{subcaption}
\usepackage{booktabs}
\usepackage{tabularx}
\usepackage{array}
\AtBeginDocument{%
  }

\setcopyright{acmlicensed}
\copyrightyear{2018}
\acmYear{2018}
\acmDOI{XXXXXXX.XXXXXXX}
\acmConference[Conference acronym 'XX]{Make sure to enter the correct
  conference title from your rights confirmation email}{June 03--05,
  2018}{Woodstock, NY}
\acmISBN{978-1-4503-XXXX-X/2018/06}

\begin{document}

\title{AuraDesk: Data Physicalization through Olfaction Metaphors for Representing and Mitigating Workplace Stress}

\author{Siying Hu}
\affiliation{%
  \institution{The University of Queensland}
  \city{Brisbane}
  \state{Queensland}
  \country{Australia}
}
\email{siying.hu@uqconnect.edu.au}
\orcid{0000-0002-3824-2801}

\author{Zhenhao Zhang}
\affiliation{%
  \institution{University of Hong Kong}
  \country{Hong Kong SAR}
  }
\email{zhenhao.research@gmail.com}
\orcid{0009-0001-9242-9618}

\renewcommand{\shortauthors}{Trovato et al.}

\begin{abstract}

  Workplace stress is often addressed through visual or auditory interventions, yet these modalities can compete with attention and contribute to sensory overload. We explore olfaction as an alternative ambient medium for representing stress-related physiological signals in office settings. We present AuraDesk, an olfactory data physicalization system that translates wearable-derived physiological cues into situated scent expressions at the workstation. The system combines local physiological state inference with a constrained actuation strategy to produce temporally regulated and spatially localized scent output suitable for everyday work environments. To examine the feasibility and experiential qualities of this approach, we conducted a one-day in-situ field deployment with 25 knowledge workers at their actual workstations. Our findings show that participants often interpreted the scent output not as an explicit alert, but as a subtle atmospheric cue that supported momentary awareness, micro-break taking, and perceived environmental attunement. At the same time, participants raised important concerns regarding scent preference, habituation, and contextual appropriateness in shared offices. This work contributes (1) an olfactory interface for physiologically driven ambient feedback in the workplace, (2) a hybrid mapping approach for coupling continuous biosignal interpretation with constrained scent actuation, and (3) empirical insights into how workers perceive, negotiate, and appropriate ambient olfactory feedback in real office contexts. Rather than claiming therapeutic efficacy, we position AuraDesk as a probe into the design space of olfactory data physicalization for workplace wellbeing and attention-sensitive interaction.
\end{abstract}


\begin{teaserfigure}
  \includegraphics[width=\textwidth]{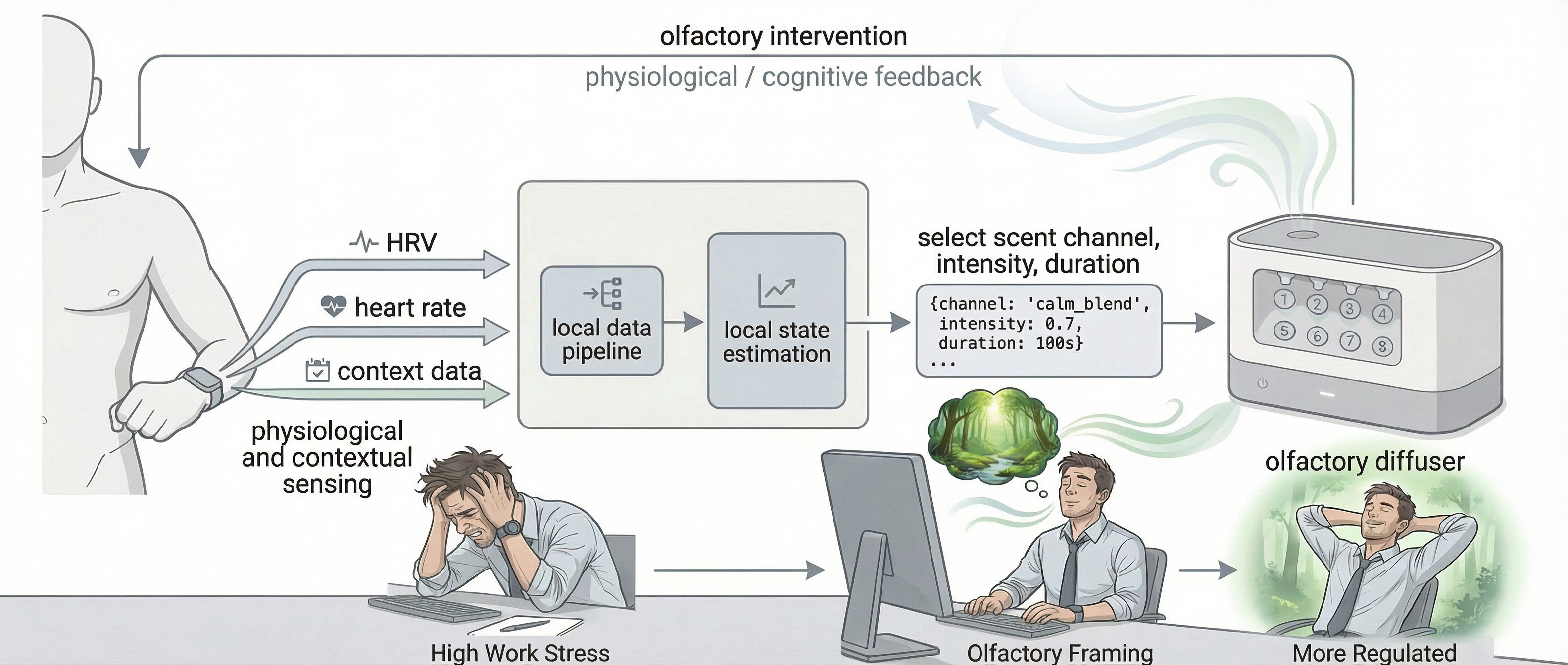}
  \caption{AuraDesk illustrates olfactory data physicalization for workplace stress during desk-based work. Physiological and contextual signals from a smartwatch are processed locally to estimate an interaction state, which is then translated into near-field scent output through the selection of scent channel, intensity, and duration. By embedding stress-related changes into the surrounding olfactory environment, AuraDesk supports low-interruption awareness and may also support regulation during ongoing work.}
  \Description{AuraDesk illustrates olfactory data physicalization for workplace stress during desk-based work. Physiological and contextual signals from a smartwatch are processed locally to estimate an interaction state, which is then translated into near-field scent output through the selection of scent channel, intensity, and duration. By embedding stress-related changes into the surrounding olfactory environment, AuraDesk supports low-interruption awareness and may also support regulation during ongoing work.}
  \label{fig:teaser}
\end{teaserfigure}

\received{20 February 2007}
\received[revised]{12 March 2009}
\received[accepted]{5 June 2009}

\maketitle

\section{Introduction}
A growing range of health and wellbeing technologies has been developed to help people monitor stress, attention, fatigue, and other aspects of mental state in everyday life \cite{a18070419, Xu2025ASA, KAKHI2025110461}. These technologies include wearable sensors and smartwatches that present heart rate variability, stress scores, and recovery indicators through dashboards \cite{info:doi/10.2196/60708, Pozzato2025WearableBased}, mobile apps \cite{Jerath2023Future, info:doi/10.2196/68012}, and notifications \cite{10.1145/3706598.3713802, Alavi2022Realtime}. They also include workplace and home-based systems that provide interventions through guided breathing \cite{HONINX2023395}, calming media \cite{Xue2025GlowGrow}, and ambient feedback \cite{10.1145/3563657.3596096} such as light or sound cues. Such systems have largely been designed around visual and auditory interaction paradigms, emphasizing metrics, alerts, and visual or auditory-based feedback as primary means for making internal states observable and actionable \cite{GonzalezRamirez2023Wearables, Jerath2023Future}.

However, these assumptions are often poorly aligned with the realities of knowledge work. In office and desk-based settings, visual and auditory channels are already saturated by screens, meetings, messages, and other ongoing demands \cite{10.3389/fpsyg.2023.1122200, MARSH2022107118}. At the same time, stress and cognitive fatigue often accumulate gradually \cite{technologies12030038, KOK2022105880}, making it difficult for workers to recognize their own deteriorating condition before exhaustion, overload, or disengagement has already set in \cite{Karakolias2025Seeing, Zhang2025Navigating}. This raises questions about the suitability of dashboards and notifications for supporting stress awareness in contexts where attention is already fragmented and interruptions are costly. For workplace stress in particular, the gap between sensed physiological change and subtle, immediately meaningful cues in everyday work remains underexplored.

To address this gap, this research explores olfactory data physicalization for implicit stress awareness and mitigation in the workplace. We present AuraDesk, an ambient interface that continuously maps physiological indicators of cognitive load and accumulated stress onto dynamic scent parameters in the surrounding workspace. Rather than asking users to inspect charts or respond to alerts, AuraDesk renders stress data perceptible through gradual changes in scent intensity, fragrance profile, and release rhythm. In doing so, it shifts interaction away from explicit self-monitoring and toward a more situated and low-interruption form of engagement with bodily data.

AuraDesk is motivated by two complementary opportunities in prior HCI research. First, ambient interfaces have shown how information can be conveyed in the periphery of attention, allowing users to remain aware without disrupting ongoing activity \cite{weiser_brown_1995_designing_calm_technology, 10.5555/1466595.1466602, hausen:hal-01513900}. Second, data physicalization has demonstrated how abstract information can be translated into material and experiential forms that support interpretation, reflection, and emotional engagement \cite{10.1145/3617366, 10.1145/3772318.3791978}. Yet most data physicalization systems have relied on visual or tangible artifacts \cite{Burzio2024Clarifying, Dumicic2022}, while most olfactory systems have used scent as a therapeutic or atmospheric intervention \cite{10.1145/3470975, 10.3389/frvir.2024.1252539, 10.1145/2816454} rather than as a medium for representing personal data. As a result, little research has examined how scent might function simultaneously as a carrier of physiological information and as a gentle regulator of affect in everyday workplace environments.

Rather than evaluating scent as a clinically validated stress intervention, we position AuraDesk as an exploration of how stress-related physiological signals can be externalized through ambient olfactory feedback in real workplace settings. Our research sought to understand:

\begin{itemize}
    \item \textbf{RQ1}: How can physiological indicators of cognitive load and accumulated stress be translated into ambient olfactory forms that remain perceptible without interrupting ongoing work?
    \item \textbf{RQ2}: How can olfactory data physicalization differ from conventional visual-based stress interfaces in supporting both awareness and emotional regulation?
\end{itemize}


Guided by these questions, this paper contributes both a system-level exploration of olfactory data physicalization and an empirical account of its use in situated workplace settings. Our contribution is threefold:
\begin{itemize}
    \item We present AuraDesk, an olfactory data physicalization system that transforms stress-related physiological signals into ambient scent expressions for office workspaces.
    \item We introduce a hybrid mapping strategy that combines physiological state inference with constrained rule-based scent actuation, addressing the temporal and spatial challenges of using olfaction as an interactive medium.
    \item Through a one-day in-situ field deployment with 25 knowledge workers at their actual desks, we provide empirical insights into how ambient olfactory feedback is interpreted, appropriated, and negotiated in real workplace settings, including its perceived benefits, limitations, and contextual sensitivities.
\end{itemize}

\section{Related Work}

\subsection{Physicalizing Personal and Bodily Data for Reflection and Awareness}
Research on data physicalization has demonstrated the value of transforming abstract data into material, embodied, and often aesthetic forms that support interpretation beyond conventional screen-based visualization \cite{Dumicic2022, Sauve25102024, 10.1145/3617366}. By externalizing data through physical properties such as shape, texture, movement, or spatial arrangement, data physicalizations can foster slower and more reflective encounters with information, inviting curiosity, interpretation, and emotional engagement \cite{10.1145/3715336.3735721, 10.1145/3689050.3704946}. Prior work has applied data physicalization across domains such as personal informatics \cite{10.1145/3623509.3635252, 10.1145/3689050.3705985}, environmental monitoring \cite{10.1145/3628516.3655788, 10.1145/3656650.3656682}, and health-related self-reflection \cite{10.1145/3772318.3791978, 10.1145/3544548.3581198}, showing how physical representations can make otherwise invisible processes more tangible in everyday life.

At the same time, this body of work reveals gaps in relation to subtle forms of personal and bodily data, such as stress, cognitive load, or fatigue. While prior data-physicalization research has predominantly emphasized artifacts encountered through vision and touch, work on these subtle states remains comparatively sparse \cite{10.1145/3563657.3596096, 10.1007/978-3-031-42280-5_26}, and in HCI more broadly, they are often assessed through retrospective reflection \cite{10.1145/3689050.3705985, 10.1145/3715336.3735721} rather than real-time physicalized representation. Consequently, existing approaches offer limited support for ongoing awareness and timely feedback during interaction \cite{10.1145/3491102.3501939, mti7070073, 10.1145/3617366}. This is particularly relevant in everyday wellbeing contexts, especially in the workplace, where users may already be visually and cognitively overloaded \cite{10.1016/j.chb.2010.03.008, 10.1145/1357054.1357072}, and where stress often accumulates gradually without entering immediate awareness \cite{10.1145/3706599.3719987, 10.1145/3544549.3585833}. As a result, little is known about how data physicalization might support awareness of ongoing bodily states in ways that are continuous, low-demand, and compatible with the background rhythms of daily life. 

\subsection{Ambient Interfaces for Peripheral Wellbeing Support}
One promising direction for addressing this limitation can be found in research on ambient and peripheral interfaces \cite{10.5555/1466595.1466602, 10.5555/645968.674740, 10.1145/985619.985617, weiser_brown_1995_designing_calm_technology}. A substantial body of HCI work has shown that ambient systems can support awareness without demanding sustained focus or interrupting primary tasks \cite{10.1145/3706598.3713511, 10.1145/3544549.3585784, 10.1145/3749507, 10.1145/3772318.3791978}. Rather than relying on dashboards, notifications, or explicit prompts, such systems communicate through subtle changes in environmental conditions, including light \cite{10.1145/3706598.3713511, 10.1145/3749507}, sound \cite{10.1145/3706598.3713786, 10.1145/3706598.3714268}, motion \cite{10.1145/3613904.3642396, 10.1145/3544548.3581486}, or material qualities \cite{10.1145/3689050.3704924}, allowing information to remain in the periphery of attention. This design approach has proven especially valuable in contexts where interruptions are costly, such as knowledge work and other cognitively intensive settings \cite{rick2024work, app8101780, 10.3389/fpsyg.2024.1465323}. Prior studies further suggest that low-salience forms of feedback can better align with the temporal and attentional rhythms of everyday routines \cite{10.1145/3491102.3517737, 10.1145/3706598.3713511}, supporting reflection while reducing the burden of active monitoring.

From this perspective, ambient interaction offers a useful lens for rethinking the translation process in data physicalization. Rather than presenting bodily data as something to be explicitly read, it may instead be embedded into the atmosphere of everyday environments \cite{10.1145/3505590, 10.1145/3491102.3501939}. This shift is particularly relevant to work on wellbeing and affective support. In parallel with developments in data physicalization, HCI researchers have explored how ambient feedback can facilitate emotion regulation and restorative experiences \cite{10.1007/978-3-031-42280-5_26, 10.1145/3544548.3581188, 10.1145/3563657.3596096} in everyday settings. However, many existing approaches continue to rely on visual or auditory modalities \cite{Shelton2020ASR}, such as color-changing displays \cite{10.1145/3706598.3713511, 10.1145/3613904.3642396}, soundscapes \cite{10.1145/3706598.3714268, 10.1145/3706598.3713786}, or notification-based prompts \cite{10.1145/3428361.3428466, 10.1145/3428361.3428400}. In workplace contexts, such channels may still compete with ongoing tasks, especially when users are already fatigued or cognitively burdened \cite{10.1145/3356590.3356601, Hinss2022Cognitive, 10.1145/3003715.3005413, 10.1145/3544549.3585784}. This points to the need for alternative ambient modalities that can remain perceptible without further taxing already saturated visual and auditory attention.

\subsection{Olfactory Interaction for Affective Support and Data Representation}
Olfactory interaction offers a particularly compelling possibility in this regard. Research on olfactory interfaces has shown that scent can shape mood, memory, attention, and the felt quality of environments in ways that differ substantially from visual and auditory media \cite{10.1145/3470975, 10.1145/1027933.1027965, MORRIN2000157}. Because olfaction is closely tied to affective processing, it has been explored as a channel for immersive interaction \cite{10.1145/3665243, 10.1145/3706599.3720290}, emotional communication \cite{10.1145/3491101.3519632, 10.1145/3025453.3026004}, environmental augmentation \cite{10.1145/2816454, 10.1145/3544549.3585832}, and therapeutic or restorative experiences \cite{10.1145/3731459.3779346, 10.1145/3025453.3026004}. Prior work also suggests that certain scents can promote calmness, reduce perceived stress, and contribute to more restorative atmospheres in workplaces \cite{Amanak2025Effects, LIU2023102135}, healthcare settings \cite{Fenko2014Influence, HEDIGAN2023101750}, and domestic environments \cite{10.3389/fpsyg.2022.791768, Herz2022ThreeFactor}. These qualities make scent especially promising as an ambient modality for wellbeing-oriented systems: it is atmospheric, can operate at the periphery of attention, and is deeply entangled with emotional experience.

Yet, despite this promise, most olfactory systems have treated scent primarily as an intervention medium rather than a representational one \cite{10.1145/3470975, 10.1145/2816454, 10.1145/3025453.3026004}. That is, scent is often used to influence users’ states, but rarely to encode and externalize those states in ways that make ongoing bodily processes perceptible. Where olfactory cues have been used for notification or contextual signaling, they have typically been event-driven \cite{10.1145/3242969.3242975, 10.1145/1027933.1027965}, symbolic \cite{10.1145/3290607.3313001, 10.1145/2775441.2775483}, or task-specific \cite{10.1145/3689050.3704927, 10.1145/3544548.3580892}, rather than continuously mapped to users’ physiological conditions. Consequently, there remains a limited understanding of how scent might function simultaneously as a carrier of data and as a medium for affective support. This gap is particularly significant in workplace settings, where stress and fatigue often accumulate gradually and are not recognized \cite{Karakolias2025Seeing, Roster2020WorkStress} until their effects on well-being or performance become difficult to ignore \cite{Habay2026MentalFatigue}. Our work addresses this gap by exploring how stress-related bodily data can be translated through ambient olfactory physicalization to support everyday wellbeing in cognitively demanding environments.

\subsection{Research gap and positioning of our work}
Prior research establishes three complementary insights. Ambient interfaces can support low interruption awareness in everyday settings. Data physicalization can make abstract and personal data more tangible and meaningful. Olfactory interaction can shape emotion and environmental experience in uniquely affective ways. Yet these strands have rarely been integrated into a single approach for representing and regulating workplace stress. Existing stress related systems still largely depend on screens, alerts, or audio visual feedback, while existing olfactory systems rarely treat scent as a dynamic physicalization of physiological data.

Our work addresses this gap through olfactory data physicalization for workplace stress and fatigue mitigation. Rather than using scent only as a calming intervention, we continuously map physiological indicators of cognitive load and accumulated stress onto olfactory parameters such as intensity, fragrance profile, and release rhythm. In this way, scent becomes both a representational medium that renders otherwise unnoticed bodily states perceptible and an affective medium that supports gentle regulation during ongoing work. This framing extends data physicalization beyond visual and tangible forms, introduces olfaction as an ambient material for embodied representations of internal state, and offers a new design direction for implicit affective interfaces in everyday workplace environments. These opportunities informed the interaction design of AuraDesk, which we describe next.

\section{Interaction Design}

Building on the gaps identified in prior work, this section presents the interaction design of AuraDesk. The design aims to support stress awareness during desk-based work with limited reliance on explicit monitoring, while using olfaction as both a representational and regulatory medium. We first outline the design goals and then describe the interaction model.

\subsection{Design Goals}

We formulated three design goals to guide the interaction design of AuraDesk.

\paragraph{\textbf{DG1. Support awareness without task interruption.}}
The system should make stress and cognitive fatigue perceptible without requiring users to inspect screens, interpret metrics, or respond to alerts during ongoing work.

\paragraph{\textbf{DG2. Represent internal state through gradual environmental change.}}
Instead of presenting stress through warnings or prompts, the system should externalize physiological change through continuous modulation of the surrounding olfactory environment.

\paragraph{\textbf{DG3. Couple representation with regulation.}}
The interaction should not only make affective state perceptible, but also use scent to support low-demand stress regulation during work.

\subsection{Interaction with AuraDesk}

Many stress monitoring systems ask users to inspect dashboards \cite{GonzalezRamirez2023Wearables, info:doi/10.2196/60708}, interpret numerical indicators \cite{10.1145/3613904.3642766, 10.1145/3613904.3642662}, or respond to alerts and prompts \cite{10.1145/3706598.3713802, Dobson2023Use}. These approaches depend on users being able to pause and attend to feedback. In desk-based knowledge work, stress and cognitive fatigue may build up while users are occupied with ongoing tasks \cite{Mahdavi2024Unraveling, Kotnik2024Prolonged}. Under such conditions, visual-based feedback may be overlooked or deferred \cite{Ohly2023Effects, 10.1145/3544549.3585784, Hinss2022Cognitive}. This suggests the need for forms of interaction that do not depend on explicit inspection.

AuraDesk addresses this issue through olfactory interaction. Rather than asking users to check stress data directly, the system maps changes in cognitive load and accumulated stress to the surrounding olfactory environment. Users may encounter this information peripherally while continuing their work. Stress is not presented as a score, chart, or warning. It is expressed as gradual atmospheric change.

AuraDesk uses three olfactory parameters to structure this interaction over time. Scent intensity reflects the level of current load. Fragrance profile represents qualitative differences in the sensed condition. Release rhythm reflects temporal variation, including short fluctuations and more sustained strain. Together, these parameters provide an olfactory representation that changes over time rather than through discrete events.

The interaction is organized as a continuous loop of sensing, modulation, and perception. As users work, the system updates scent conditions in response to stress-related physiological signals. During periods of lower strain, the olfactory output remains less salient. As strain increases, changes in intensity, profile, or rhythm may become more noticeable. This may support awareness without requiring immediate action.

The same interaction channel is also used for regulation. When the system detects sustained or elevated stress, it may shift the olfactory output toward calmer scent qualities rather than only increasing representational salience. In this way, awareness and regulation are integrated within the same interaction. The changing scent both reflects sensed condition and may support stress mitigation.

Compared with conventional biofeedback systems, this interaction model reduces reliance on screen-based display and discrete notifications. It also treats olfaction as a medium for representing physiological change, not only as an added intervention. Taken together, this design explores how workplace stress may be externalized and potentially regulated through ambient olfactory data physicalization.

\section{System Design and Implementation}
AuraDesk is a desk-side olfactory interface that translates physiological and contextual signals into sparse, near-field scent cues during office work. The prototype comprises three functional layers: wearable and contextual sensing, local state estimation and output scheduling, and multichannel olfactory actuation. During desk-based work, the user wears an smart watch, while an NVIDIA Jetson Nano (P3450) serves as the local control hub for signal aggregation, lightweight state estimation, and actuation scheduling. An eight-channel olfactory device positioned near the workstation delivers short scent releases close to the user rather than diffusing fragrance at room scale. This design emphasizes bounded intervention, local control, and reproducible desk-side deployment.

\begin{figure*}
    \centering
    \includegraphics[width=1\linewidth]{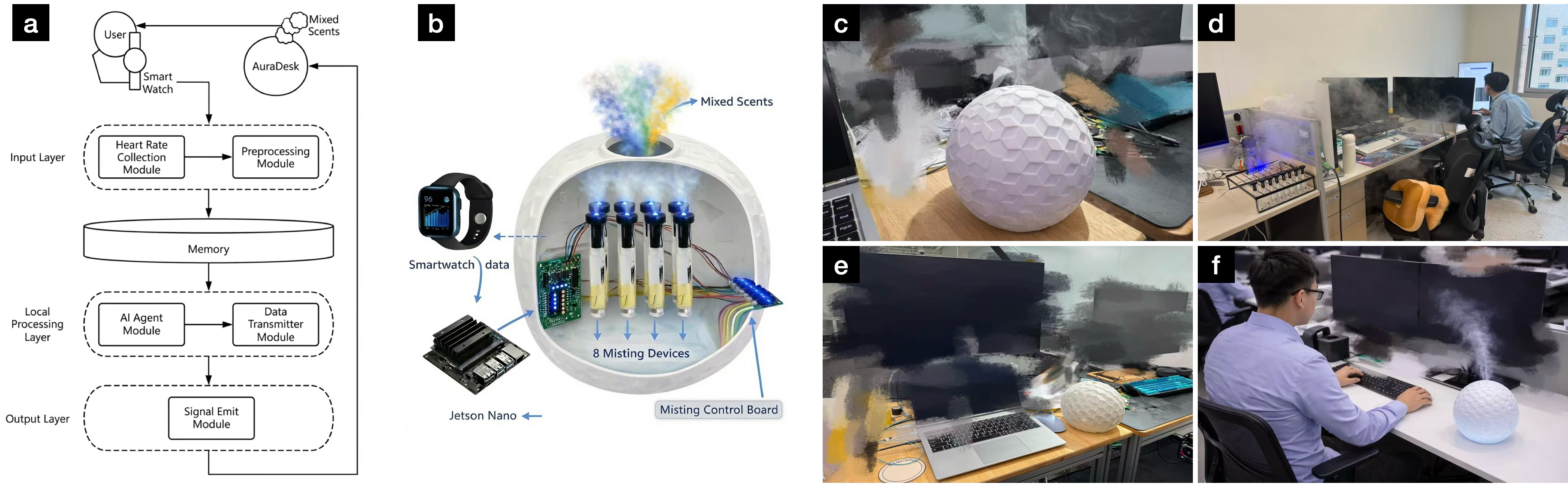}
    \caption{The overview of schematic diagram: (a) Processing map of the AuraDesk system. (b) Internal structure of the AuraDesk device. (c) AuraDesk in operation with blue ambient lighting and mixed-scent output. (d) Pilot study with low-fidelity prototype. (e) The office-environment setup of AuraDesk. and (f) User interaction with AuraDesk in a real office scenario.}
    \label{fig:auradeskoverview}
\end{figure*}

\subsection{Scent Selection through Olfactory Metaphor}
AuraDesk organizes its scent vocabulary through olfactory metaphor. Scent is not used as a direct encoding of discrete emotions. Instead, it is used as an ambient cue \cite{MORRIN2000157, 10.1145/1027933.1027965} for familiar environmental scenes whose atmospheric qualities can support regulation during desk based work. Rather than assigning each scent a fixed semantic meaning, the system uses scent cues to suggest scene like impressions such as forest, sea breeze, grassland, and garden. Users can perceive these cues peripherally and interpret them in context. This approach is appropriate for olfaction because scent is commonly experienced as an associative and atmospheric medium rather than as an explicit symbolic signal.

Within this vocabulary, woody and resinous notes such as cedarwood \cite{Zhang2018Juniperus}, frankincense \cite{Okano2019Frankincense}, and vetiver \cite{Saiyudthong2015Vetiver} support forest related metaphors associated with grounding, steadiness, and depth. These notes are suitable for calming or stabilizing moments. Fresh herbal notes such as peppermint \cite{Ilmberger2001Alertness} and tea tree \cite{VORA2024118180} suggest open air, sea breeze, or rain cleared surroundings. These notes support attentional reset, cleansing, and renewed alertness. Citrus and floral green notes such as bergamot \cite{LIU2023102135}, litsea cubeba \cite{Chen2012Litsea}, and rose geranium \cite{RashidiFakari2015Geranium} suggest brighter scenes such as garden, grove, or grassland. These notes support pleasantness, balance, and gentle activation. These scene metaphors are not intended to reproduce real environments. They provide a compact structure for the output vocabulary and help users form consistent experiential associations.

To implement this strategy, we selected eight single note essential oils from distinct scent families. Single note oils were preferred over blended scents because they preserve clearer olfactory identity, simplify routing logic, and make scene associations easier to interpret and reproduce. In the current prototype, these oils support five broad interaction roles, calming, energizing, balancing, pleasantness, and cleansing, while remaining grounded in a smaller set of scene oriented metaphors summarized in Table~\ref{tab:scent_vocabulary}. The design goal is not to communicate a literal message through scent. The goal is to introduce a bounded environmental suggestion that can modulate the experiential quality of the workspace.

\begin{table*}[t]
\centering
\small 
\caption{Scent vocabulary, scene metaphors, and intended interaction roles in AuraDesk. Rather than mapping scents to rigid emotion labels, AuraDesk organizes its olfactory vocabulary through scene-oriented metaphors such as forest, sea-breeze-like openness, grassland, and garden. These metaphors describe the atmospheric character each scent is intended to suggest, allowing the system to support regulation through environmental implication rather than explicit symbolic messaging.}
\label{tab:scent_vocabulary}
\begin{tabularx}{\textwidth}{l l l X l}
\toprule
Scent name & Scent family & Scene metaphor & Atmospheric quality & Primary interaction role \\
\midrule
Bergamot & Citrus & Sunlit grove / orchard edge & Bright, pleasant, lightly uplifting & Pleasantness \\ 
Rose geranium & Floral-green & Garden / soft grassland bloom & Soft, rounded, gently balancing & Balancing / Pleasantness \\
Peppermint & Herbal & Cool open air / sea-breeze-like freshness & Fresh, crisp, activating & Energizing \\
Tea tree & Leafy-herbal & Rain-cleared air / coastal freshness & Clean, clarifying, resetting & Cleansing \\
Himalayan cedarwood & Woody & Forest / wooded shelter & Grounding, warm, steady & Calming \\
Frankincense & Resinous-balsamic & Quiet forest interior / still sanctuary & Centering, composed, reflective & Calming / Balancing \\
Vetiver & Rooty-earthy & Forest floor / deep terrain & Deep, anchoring, stabilizing & Calming \\
Litsea cubeba & Seed profile & Bright meadow / citrus field edge & Vivid, lively, gently stimulating & Energizing / Pleasantness \\
\bottomrule
\end{tabularx}
\end{table*}

\subsection{Mapping Strategy}
\begin{figure}
    \centering
    \includegraphics[width=1\linewidth]{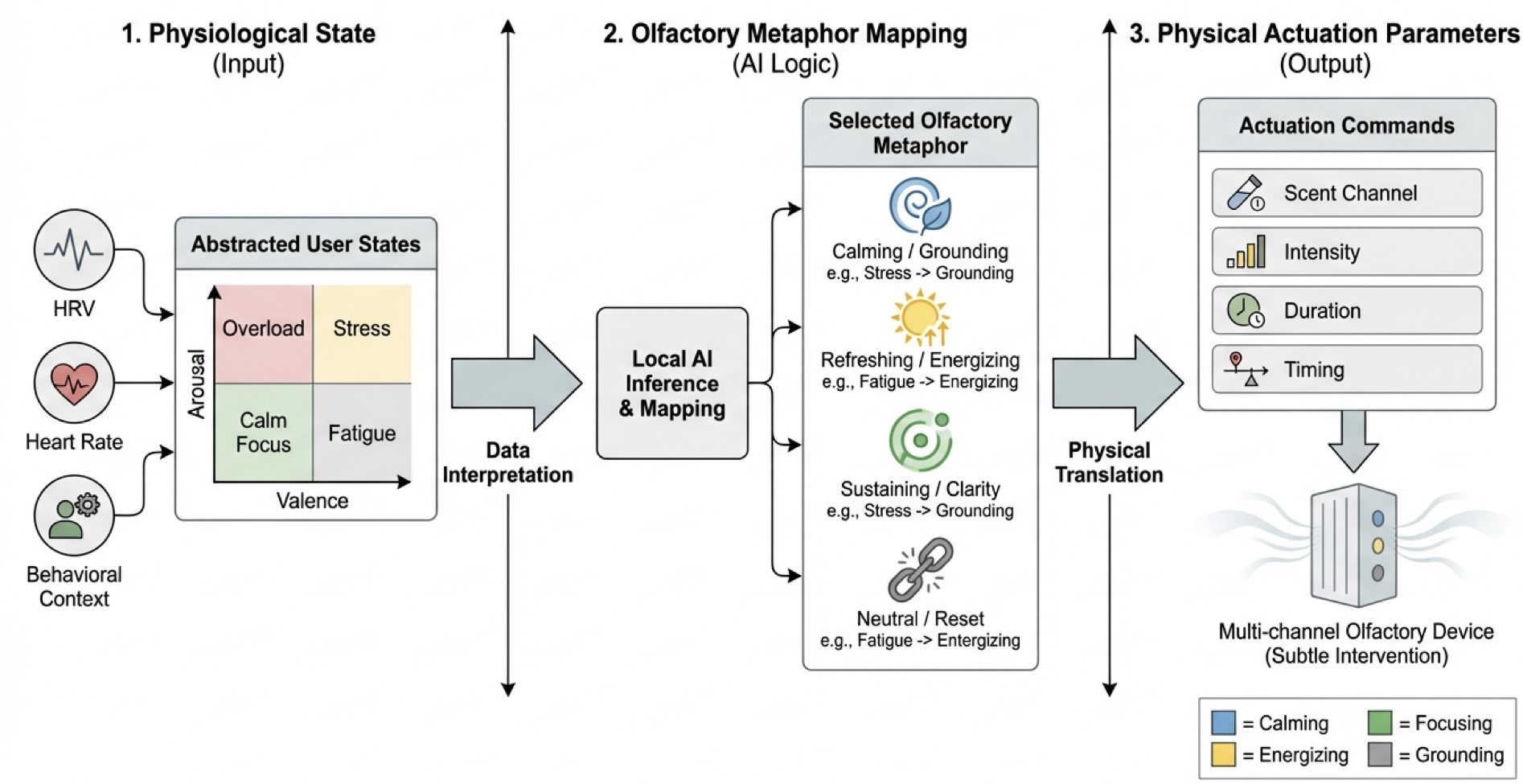}
    \caption{The Olfaction Metaphors Logic. This matrix illustrates the mapping between physiological data states (derived from HRV and HR) and specific olfactory interventions. PicoLM interprets the user's position within the Arousal-Valence space and triggers the corresponding scent metaphor (e.g., Peppermint for fatigue, Cedarwood for stress) to gently nudge the user towards an optimal cognitive state.}
    \label{fig:MetaphorsLogic}
    \vspace{-0.5em}
\end{figure}

AuraDesk uses a hybrid neuro-symbolic mapping strategy to translate noisy physiological inputs into stable olfactory outputs \cite{ALI2015601, Sattayakhom2023ScopingReview}. In the first stage, a lightweight local AI model, PicoLM, processes HRV-centered physiological signals and contextual cues. PicoLM maps these multidimensional time-series inputs into a continuous interaction state within the Arousal-Valence space (as illustrated in \autoref{fig:MetaphorsLogic}). In the second stage, to bridge the gap between fast-changing AI inferences and the slow-dissipating nature of olfactory actuation, we implement a constrained rule-based scheduling layer. Rather than allowing the AI to continuously trigger the hardware, this rule set discretizes the PicoLM's Arousal-Valence coordinates \cite{Russell1980Circumplex} into specific scene-oriented olfactory expressions (profile, intensity, and rhythm). This deterministic layer favors stability, enforces minimum inter-release intervals, and prevents olfactory fatigue or scent mixing. \autoref{tab:mapping_strategy} summarizes these rule-based state-to-output mappings.

\begin{table*}[t]
\centering
\small
\caption{Representative state to output mappings in AuraDesk. Interaction states are translated into scene oriented olfactory expressions defined by profile, intensity, and rhythm.}
\label{tab:mapping_strategy}
\begin{tabular}{@{} >{\raggedright\arraybackslash}p{0.17\textwidth} >{\raggedright\arraybackslash}p{0.27\textwidth} >{\raggedright\arraybackslash}p{0.11\textwidth} >{\raggedright\arraybackslash}p{0.15\textwidth} >{\raggedright\arraybackslash}p{0.24\textwidth} @{}}
\toprule
Interaction state & Profile & Intensity & Rhythm & Intended effect \\
\midrule
Elevated stress with persistence
& Forest profile using cedarwood, frankincense, or vetiver
& Medium to high
& Repeated low frequency release
& Support calming and stabilization during sustained strain \\
Elevated stress with short duration
& Forest profile using cedarwood or frankincense
& Low to medium
& Brief single release
& Provide a light intervention without overreacting to transient change \\
Recovery after prior strain
& Garden or grove profile using rose geranium or bergamot
& Low
& Brief single release
& Support rebalancing and ease the transition back to a neutral state \\
Low alertness during continuous work
& Sea breeze or open air profile using peppermint or tea tree
& Medium
& Brief release, repeat only if needed
& Promote attentional reset and renewed alertness \\
Mild imbalance without clear stress escalation
& Garden or grassland profile using bergamot, rose geranium, or litsea cubeba
& Low
& Brief single release
& Introduce pleasantness or balance through a light environmental cue \\
\bottomrule
\end{tabular}
\end{table*}

\subsection{System Architecture and Implementation}
AuraDesk uses three layers for desk based olfactory interaction, including physiological sensing, local state estimation, and near field scent output. Preprocessing and feature construction are shown in \autoref{fig:auradeskoverview}.

\paragraph{\textbf{Input layer.}}
AuraDesk uses heart rate variability from a smart watch\footnote{HUAWEI WATCH 5, https://consumer.huawei.com/en/wearables/watch-5/specs/} as its main physiological input. To improve robustness, HRV is interpreted together with lightweight contextual cues, including heart rate trend, activity state, and work session continuity when available. The system is not intended for clinical assessment. It estimates a compact interaction state for ambient olfactory output. Because physiological signals vary at shorter timescales than scent output, incoming measurements are temporally aggregated and smoothed before state estimation. This reduces sensitivity to short fluctuations and better matches the slower onset and dissipation of scent. 

\paragraph{\textbf{Local processing layer.}}
An NVIDIA Jetson Nano\footnote{NVIDIA Jetson Nano (P3450), https://developer.nvidia.com/embedded/jetson-nano} serves as the local edge controller. It runs the quantized PicoLM\footnote{PICOLM, https://github.com/RightNow-AI/picolm} model locally to ensure privacy and low-latency state estimation without relying on cloud transmission. The Jetson Nano also executes the rule-based actuation logic, scheduling channel selection, release duration, and inter-release intervals based on the AI's output. The implementation uses a Jetson Nano P3450 core module and carrier board with a 128\,GB microSD card for the Ubuntu system image, deployed model, and automation scripts. In the control path, the Jetson Nano sends pulse signals through its GPIO pins to an infrared transmitter module, which converts them into a 38\,kHz infrared signal received by the atomization control board and mapped to the corresponding actuation command.

\paragraph{\textbf{Output layer and hardware implementation.}}
As shown in \autoref{fig:auradeskoverview}, AuraDesk delivers scent through a custom desk side assembly built around an eight channel USB atomization control board with infrared remote functionality. Each channel contains a water based atomization medium infused with a single note essential oil. To reduce carryover, each channel uses an independent short outlet path that terminates near the top of the device rather than entering a shared diffusion chamber. The olfactory layer uses eight microporous atomization discs operating at 108 to 110\,kHz, with one absorbent cotton wick per channel. Infrared control uses an amplified transmitter module, since the Jetson Nano GPIO pins cannot reliably drive a bare infrared LED. Command transmission is based on codes recorded from the bundled remote control, including power, shutdown, and channel selection commands for channels 1 to 8. Code acquisition can be performed with a receiver module.

The enclosure includes a 5\,V silent fan mounted near the bottom to improve airflow and help clear residual mist. Power is supplied to the Jetson Nano through a 5\,V/4\,A DC adapter via the barrel jack. Standard Dupont jumper wires are used for connections among the GPIO pins, infrared modules, and fan. For desk deployment, AuraDesk emits short localized bursts near the workstation rather than room scale diffusion. Only one channel is activated at a time, and a minimum interval is enforced between releases. These constraints help reduce scent mixing, olfactory fatigue, and accumulation in shared settings. Because scent has a slower onset and dissipation profile than visual or auditory feedback, the system uses sparse scent events rather than continuous modulation or rapid switching. The control logic varies profile, intensity, and rhythm through channel selection, release duration, and inter release timing. Table~\ref{tab:deployment_constraints} summarizes the main deployment constraints and their implications for system behavior.

\begin{table*}[t]
\centering
\small
\caption{Key deployment constraints and their implications for olfactory output behavior in AuraDesk.}
\label{tab:deployment_constraints}
\begin{tabular}{@{} >{\raggedright\arraybackslash}p{0.3\textwidth} >{\raggedright\arraybackslash}p{0.3\textwidth} >{\raggedright\arraybackslash}p{0.35\textwidth} @{}}
\toprule
Constraint & Rationale & Design Implication \\
\midrule
One active channel at a time
& Reduces overlap between scents and preserves clarity of each release
& Output is scheduled as discrete scent events rather than combined releases \\
Minimum interval between releases
& Limits olfactory fatigue and prevents excessive accumulation
& The system favors sparse actuation and avoids rapid switching \\
Short localized bursts near the workstation
& Supports desk use while reducing spread in shared space
& Output remains near field, bounded, and compatible with routine office deployment \\
\bottomrule
\end{tabular}
\end{table*}

\vspace{-0.5em}
\section{Study Method}
We conducted an exploratory study to examine how physiological indicators of cognitive load and accumulated stress could be translated into ambient olfactory cues during desk-based work. Rather than evaluating clinical efficacy or testing whether AuraDesk improved productivity or reduced stress in a causal sense, the study focused on how participants perceived, interpreted, and responded to state-driven scent interventions in a workplace-like setting. In particular, we investigated whether such cues could remain perceptible without interrupting ongoing work, and how participants experienced this modality in relation to more familiar visual- or auditory-based stress interfaces.

Methodologically, the study adopted a mixed-methods approach, combining system logs, post-study ratings, and qualitative interviews. However, our emphasis was primarily qualitative and interpretive. This orientation was appropriate because our research questions concerned how participants made sense of olfactory representations of physiological state, what forms of scent delivery they found acceptable or meaningful, and how they distinguished this modality from more conventional interfaces. Accordingly, quantitative measures were used to provide contextual grounding and to support triangulation, rather than to test effectiveness or support strong causal claims.

\subsection{Participants}
We recruited 25 adults (aged between 18 and 30, M = 28) who regularly engaged in desk-based computer work. We focused on desk workers because AuraDesk is intended for deployment at individual workstations, where olfactory output must remain subtle, localized, and compatible with everyday work routines. This population was also appropriate for our research questions, as desk-based work often involves fluctuating workload, transient stress, and attentional fatigue over sustained periods of computer use.

Participants had normal or corrected-to-normal vision and reported no acute respiratory illness, severe fragrance sensitivity, or known olfactory impairment. To reduce confounds in scent perception, participants were asked to avoid strong perfume or heavily scented personal care products on the study day. All participants provided informed consent prior to participation. We additionally collected demographic and contextual information, including age, gender, prior experience with scented products, and self-reported smell sensitivity.

\subsection{Procedure}
Each participant completed a one-day session at their own desk in their regular office, temporarily equipped with AuraDesk. We adopted this in-situ protocol to balance ecological validity with logistical feasibility. This deployment captured genuine short-term fluctuations in workload, stress, and fatigue amidst actual daily tasks and natural work rhythms, while still allowing us to control baseline calibration and scent exposure parameters.

Each session comprised four phases: orientation, baseline calibration, task engagement, and post-study reflection. During orientation, participants were introduced to the setup and acclimated. During baseline calibration, participants sat quietly to establish an individual reference state. These baseline measurements were later used to interpret within-person deviations during the task phase.

The task phase involved desk-based activities approximating sustained office work. We used a structured work--rest schedule informed by ergonomics research showing mental fatigue emerges after 30--45 minutes of continuous work, and 5--10 minute breaks reduce cognitive strain. In our study, breaks also helped reduce olfactory adaptation and scent carryover across releases.

AuraDesk monitored user state continuously and delivered olfactory output according to a predefined state-to-scent mapping. To avoid overlap and sensory saturation, only one scent channel was active at a time, with enforced intervals between releases. Because stress and fatigue responses vary across individuals, the system relied on participant-relative deviations from baseline rather than fixed population-wide thresholds. Participants were not interrupted during routine operation except at predefined measurement points.

After the task phase, participants completed a post-study survey and a semi-structured interview. The survey asked participants to rate several key experiential dimensions of AuraDesk, including the noticeability of scent cues, their comfort and pleasantness, timing appropriateness, non-intrusiveness during work, interpretability of the physiological-state-to-scent mapping, and perceived usefulness for supporting awareness and regulation. Survey items were rated on a 7-point Likert scale ranging from 1 (strongly disagree) to 7 (strongly agree). These questions were designed to address RQ1, namely whether stress-related physiological changes could be translated into ambient olfactory forms that remained perceptible without interrupting ongoing work, and RQ2, namely how this olfactory modality differed from more conventional visual-based stress interfaces (details in \autoref{appendix:survey-items}). We then conducted a semi-structured interview to further examine how participants noticed and interpreted the scent cues, how they experienced the atmospheric and emotional qualities of the intervention, and how they compared this modality with more familiar visual- or auditory-based forms of stress feedback. This comparative component was reflective rather than experimental: we did not run a separate controlled screen-based condition, but instead used participants' accounts to understand modality-specific differences in attentional demand, interpretability, and perceived emotional effect.

\subsection{Data Collection and Analysis}
We collected system logs, post-study survey ratings, and qualitative interview data. System logs recorded the timing, profile, intensity, and sequencing of scent releases. The post-study survey captured experiential dimensions including comfort, noticeability, pleasantness, timing appropriateness, non-intrusiveness, interpretability of the scent-state mapping, and perceived usefulness during work. These items were intended to assess whether olfactory cues could remain perceptible without interrupting ongoing activity, and how participants experienced this modality in relation to more familiar visual or auditory stress feedback.

Semi-structured interviews examined how participants made sense of the physiological state-to-scent mapping, how they experienced the cues while working, and how they compared AuraDesk with conventional interfaces. The interview also invited participants to reflect on the broader meaning of the experience, including how scent shaped their awareness of stress and fatigue, its influence on emotional states, and the role of olfactory cues in the workspace atmosphere.

Our analysis was primarily qualitative and interpretive. We used descriptive statistics to summarize survey ratings and system logs, and thematic analysis to identify recurring experiential patterns across participants. Quantitative measures were included to contextualize participants' experiences, characterize the session conditions, and triangulate emerging themes; they were not intended to establish efficacy, compare modalities experimentally, or demonstrate that olfactory feedback produced measurable stress reduction.

\section{Findings}
\begin{figure}
    \centering
    \includegraphics[width=0.9\linewidth]{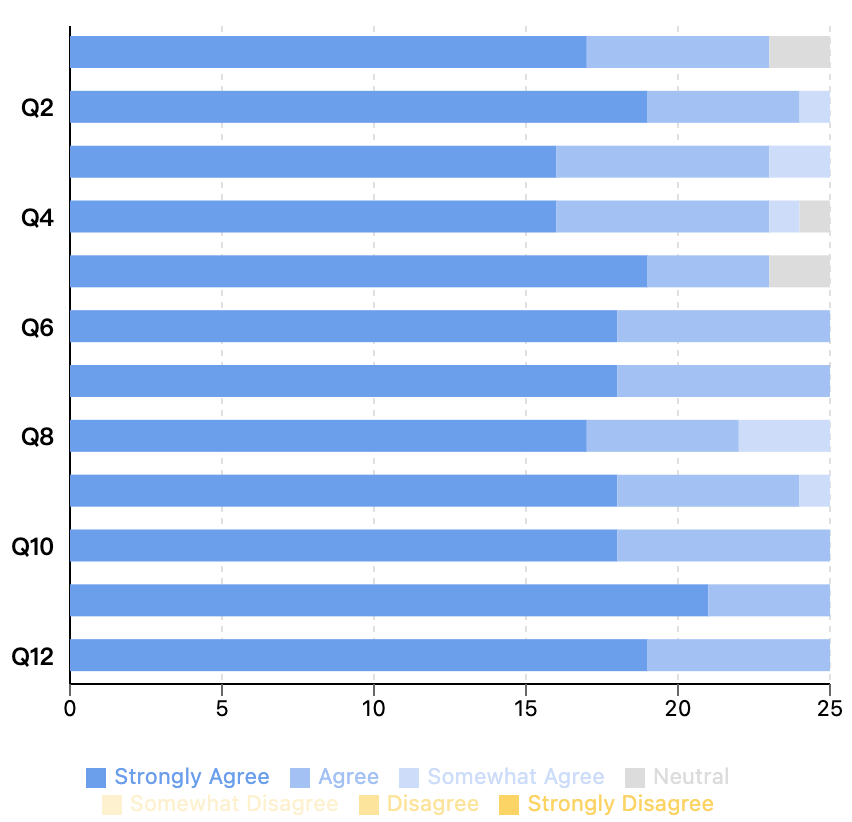}
    \caption{Results of the post-study questionnaire. Participants rated the noticeability, comfort, and overall usefulness of AuraDesk's scent cues, indicating highly positive feedback across all items.}
    \label{fig:survey}
    \vspace*{-10pt}
\end{figure}

Our analysis shows that participants did not frame olfactory feedback as a direct equivalent of visual stress displays. Instead, they imagined scent as an ambient, affective, and context-sensitive form of data physicalization. In discussing how physiological indicators of cognitive load and accumulated stress might be represented through smell, participants emphasized environmental modulation, subtle delivery, dynamic variation, and personally meaningful associations. In general, participants found AuraDesk’s scent cues noticeable, understandable, and supportive of awareness and regulation, as shown in \autoref{fig:survey}. 
We organize the findings into two subsections corresponding to our two research questions.

\subsection{RQ1: Translating physiological state into ambient olfactory forms}
In this subsection, we showed how stress-related physiological states may be mapped into olfactory output. Participants consistently suggested that such translation should work through atmospheric change rather than explicit notification, and that its effectiveness depends on subtle diffusion, controllable boundaries, and temporal variation.

\subsubsection{Physiological stress was more acceptable when translated into atmospheric change rather than explicit olfactory notification}
A central finding is that the participant did not imagine scent as a discrete, alert-like output that explicitly announces stress or fatigue. Instead, they repeatedly reframed the intervention as a modification of the surrounding atmosphere. 
As P3 noted, \textit{If it's not just to perfume, but to change the scent of the environment, I think it's quite nice, and it makes it easier for us to maybe associate it with the smell of the place that I was in, the smell of the place that I stayed in.''} This suggests that physiological signals may be more appropriately translated into olfactory forms when the system subtly alters the felt quality of the environment, rather than introducing an abrupt, message-like cue. This preference for an unobtrusive ambient shift is further supported by P12, who appreciated the system's subtlety, stating: \textit{It doesn't suddenly irritate you, it's gentle and you can take it very gently.''}
In this sense, olfactory representation appears to function less as notification and more as atmospheric modulation.

\subsubsection{Perceptibility without interruption depended on subtle intensity, bounded diffusion, and controlled presence}
When discussing use in shared settings, the participant did not simply say that the scent should be noticeable. The interview suggests that non-interruptive olfactory feedback is not simply a matter of making scent detectable. Instead, perceptibility should be balanced against the risk of becoming socially intrusive. The participant emphasized that scent release should be low in intensity, spatially limited, and deployable in contexts where exposure can be controlled. 
As Participant P3 noted: \textit{``It doesn't feel intrusive or out of place in an office environment, but the amount of a single spray shouldn't be too much, as I don't want to bother others. It's hard to guarantee that everyone will like it. Since some people might not necessarily appreciate the scent, I feel it could sometimes cause inconvenience. Therefore, I prefer to use it in situations that are under my control.''}
This indicates that the translation of physiological stress into olfactory form depends not only on sensory thresholds but also on social acceptability. Ambient olfactory feedback should therefore remain noticeable to the user while minimizing unwanted spillover to others.

\subsubsection{Translating stress into scent required dynamic variation rather than fixed one-to-one mapping}
Rather than proposing a fixed mapping between a particular stress state and a single scent, the participant described an olfactory system based on variation, recombination, and changing proportions. Our findings indicate that users highly value dynamic scent generation over a single, static fragrance to maintain its relaxing effect. 
As P7 suggested: \textit{``Could the scent be made more variable? ... Instead of spraying a single type of perfume, it could be a blend of different elements, allowing me to experience a slightly different scent each time. Rather than a specific scent I already know I like, a novel but not unpleasant aroma might actually bring me a greater sense of relaxation.''} This preference for novelty is closely linked to the need to prevent olfactory adaptation. A similar sentiment was shared by P11, who noted during the post-experiment feedback that a repetitive scent would lose its efficacy in alleviating fatigue: \textit{``I probably wouldn't want to smell the exact same scent ten times... If I smell the same scent all ten times I feel tired, I would probably still feel tired.''} Together, these insights highlight the importance of introducing variability in scent-based interventions.
This suggests that translating physiological stress into scent may require dynamic modulation rather than stable one-to-one encoding. Repetition was seen as diminishing effectiveness; if the same smell appeared every time fatigue occurred, it could lose its refreshing quality or even become associated with tiredness itself. Sustained usefulness therefore seems to depend on temporal variability as much as on perceptual noticeability.

\subsection{RQ2: Differences from visual-based stress interfaces}
In the section, we showed how the participant understood olfactory physicalization as distinct from more conventional screen-based or explicitly informational interfaces. Rather than valuing scent for standardized readability, the participant described it as an environmental medium that could shift mood, evoke imagined places, and derive meaning from personal memory and preference.

\subsubsection{Olfactory cues supported awareness at the periphery of attention rather than through explicit monitoring}
The participant’s account suggests that olfactory feedback was appealing precisely because it functioned as a background environmental shift rather than an object of focused attention. Unlike visual representations that often require users to look, read, and interpret, scent here was imagined as something sensed in passing through the atmosphere of the workspace. 
Participants repeatedly described the system's effect as \textit{``changing the scent of the environment (P3)''} or \textit{``refreshing the ambient smell (P7),''} positioning the olfactory output as an environmental modification rather than a direct alert. Reiterating this core function, P12 noted, \textit{``Its sole purpose is to change the ambient scent; anything different from the current stuffy office smell will do.''} To illustrate this spatial transformation, P22 suggested concrete thematic applications: \textit{``Could we set up specific modes, for example, a seaside, grassland, or snowfield, to generate a related outdoor fragrance?''}
Olfactory physicalization therefore appears to support awareness at the periphery of attention, not through explicit monitoring but through embodied perception of contextual change.

\subsubsection{Olfactory physicalization was valued for coupling awareness with emotional regulation}
A major distinction is that the participant did not treat scent as merely another channel for communicating stress information. Instead, they saw it as a medium that could simultaneously make them aware of their condition and shift their emotional state. 
P3 also noted that the system serves as an effective coping mechanism, stating: \textit{``It can be used in this way to alleviate fatigue or help regulate my mood, especially when I'm feeling sleepy, or when I'm extremely frustrated but have too much work to actually leave my desk.''} This desire for a brief mental escape from the workstation was echoed by P17, who explained how the scent facilitates spatial imagination: \textit{``Because the air feels so fresh, it makes it easier for us to mentally connect with those places we wish we could go to but physically cannot.''}
The imagined value of scent lay in its ability to produce relaxation, novelty, and a sense of mental relocation rather than simply deliver an interpretable signal. Compared with conventional visual interfaces, which often foreground representation and require subsequent self-regulation, olfactory physicalization appears to integrate awareness and affective modulation within the same sensory experience.

\subsubsection{The effectiveness of olfactory physicalization depended on personally meaningful associations rather than universal categories}
The participant repeatedly emphasized that the value of scent depended on whether it felt personally acceptable, emotionally resonant, and experientially meaningful. 
To illustrate this preference for novel environmental scents, one participant provided a specific example: \textit{``I would choose a novel but not unpleasant scent, which might bring me a greater sense of relaxation. For example, I love the feeling of rain, so I would want to experience the smell right after it rains at my workstation. (P10)''} This demonstrates how specific olfactory cues can momentarily transport users away from their immediate office environment. 
This example of the smell after rain was important not because it encoded stress in a universal way, but because it carried a remembered sense of freshness. This suggests that olfactory physicalization differs from many visual interfaces, which often rely on standardized legibility across users. Here, effectiveness depended less on universal symbolic clarity and more on personal memory, affective resonance, and individual preference.

\section{Discussion}
Our study aimed to explore how workers interpret, imagine, and negotiate ambient olfactory representations of physiological states. 
The results show that ambient olfactory physicalization should not be understood as a simple sensory substitute for visual stress displays. \cite{10.1145/3505590, 10.1145/3470975, 10.1145/1027933.1027965, MORRIN2000157} For RQ1, participants indicate that physiological stress is more appropriately translated into subtle, bounded, and dynamically varying atmospheric cues. For RQ2, olfactory physicalization differs from traditional visual interfaces in that it supports peripheral awareness, couples sensing with emotional regulation, and derives meaning through personal associations rather than standardized representations.

\subsection{Rethinking Stress Representation through Olfactory Physicalization}
In light of our results, it may be useful to shift workplace stress tracking from explicit monitoring \cite{10.1145/3706598.3713802, Dobson2023Use} to more suggestive forms of representation. Because the system did not provide precise stress values, it reduced the sense of being judged or watched. Participants could register the cue without treating it as a demand for immediate action. Such ambiguity may also limit intervention \cite{10.5555/1466595.1466602, 10.5555/645968.674740, 10.1145/985619.985617, weiser_brown_1995_designing_calm_technology}. When a scent cue was ignored, it was difficult to determine whether this reflected user choice or limited salience of the cue itself. Instead of seeking a fixed balance between clarity and subtlety, designers may need to support negotiation and leave final control to the user. From this perspective, olfactory representations are better understood as a complementary layer of stress support than as a replacement for dashboards or other tracking tools.

\subsection{Designing Olfactory Micro-Escapes through Ambient Attunement}
Our findings suggest that olfactory physicalization can support a more ambient way of engaging with physiological stress data in workplace settings. Rather than requiring active inspection of metrics \cite{GonzalezRamirez2023Wearables, Jerath2023Future} or responding to explicit alerts \cite{10.1145/3706598.3713802, Alavi2022Realtime}, workers could encounter stress information as part of the surrounding desk environment. Scent allowed physiological change to be noticed at the periphery of attention, without fully interrupting ongoing activity \cite{weiser_brown_1995_designing_calm_technology, 10.5555/1466595.1466602, hausen:hal-01513900}. This suggests that olfactory feedback may support a form of low demand awareness that is better aligned with the attentional conditions of desk based work \cite{10.1016/j.chb.2010.03.008, 10.1145/1357054.1357072}.

This ambient mode of engagement also supported micro escapes from the workstation \cite{10.5555/1466595.1466602, 10.5555/645968.674740, 10.1145/985619.985617, weiser_brown_1995_designing_calm_technology}. Participants described these escapes in two ways. One was mental relocation. Scene based scents such as forest or sea breeze helped them connect with places they wished they could go to but physically could not (P17). The other was sensory revitalization. P4 wished for ``the smell of meals, sweet, or spicy food,'' explaining that the work environment already makes people want to escape and feels ``slightly dead, colorless, and tasteless.'' These accounts suggest that the value of scent lies not only in making stress perceptible, but also in reintroducing vitality into a sensory flat office environment.

At the same time, the same qualities that make olfactory feedback unobtrusive \cite{10.1145/3470975, 10.1145/1027933.1027965, MORRIN2000157} may also make it less explicit and less immediately actionable than visual or notification based systems. We therefore do not position olfactory physicalization as a replacement for dashboards or other self-tracking tools \cite{GonzalezRamirez2023Wearables, info:doi/10.2196/60708, 10.1145/3613904.3642766, 10.1145/3613904.3642662}. Its value lies in offering a complementary form of situated coping that links ongoing stress awareness with brief moments of restoration during everyday work.

\subsection{Designing with Olfactory Metaphors}
Our findings suggest that olfactory metaphors can contribute to experience centred design by supporting affective engagement with stress \cite{Amanak2025Effects, LIU2023102135}. Rather than presenting stress as a value to be read, olfactory physicalization connected physiological change to sensory references drawn from everyday life, memory, and imagined places. Participants interpreted scents such as forest, sea breeze, and food related smells not as neutral indicators, but as prompts for personal meaning making. In this sense, olfactory metaphors did not only encode stress. They helped users relate stress to comfort, distance, vitality, and depletion in ways that were situated and affectively meaningful.

This finding points to an alternative direction for affective interfaces \cite{10.1145/3470975, 10.1145/2816454, 10.1145/3025453.3026004}. Stress technologies are often designed for clarity, legibility, and immediate recognition \cite{10.1145/3689050.3704927, 10.1145/3544548.3580892}. Our findings suggest that some forms of ambiguity may also be useful when the goal is not only awareness, but also reflection and coping \cite{Karakolias2025Seeing, Roster2020WorkStress}. Olfactory metaphors allowed users to engage with stress through association rather than direct interpretation \cite{10.1145/3689050.3705985, 10.1145/3715336.3735721}, which may soften feelings of surveillance and make the interaction easier to accommodate in everyday work.

At the same time, the meaning of olfactory metaphors is not fixed \cite{10.1145/3470975, 10.1145/2816454, 10.1145/3025453.3026004}. It depends on personal experience, cultural context, and the situation in which the scent is encountered. A metaphor that feels grounding or restorative for one user may feel distracting or unclear for another. This suggests that designers should treat olfactory metaphors as open ended cues rather than stable symbolic mappings. Supporting personalization and revision may therefore be important when using scent to represent affective states.

\subsection{Limitations and Future Work}
This study has several limitations. First, the sample size was limited to 25 participants, which is appropriate for an exploratory study but constrains statistical power and generalizability across broader worker populations. Second, the deployment lasted only one day, allowing us to assess short-term usability and acceptance but not longer-term adaptation, habituation, or sustained behavioral effects. Third, while the study was conducted in-situ at participants' actual workstations, the deployment lasted only one day. The presence of the novel hardware and the short duration may introduce a novelty effect, limiting our ability to assess long-term ecological validity, scent habituation, or sustained behavioral changes over weeks or months. Fourth, although we incorporated quantitative elements such as system logs and post-study ratings, these were used to contextualize and triangulate the qualitative findings and to characterize participants’ experiences, rather than as a formal test of efficacy. In addition, our comparison with visual or notification-based stress interfaces was reflective rather than experimental, and should therefore be interpreted as experiential rather than performance-based.

Future work should evaluate AuraDesk with larger and more diverse participant groups and through longitudinal, in-situ deployments in real workplace settings. Such studies would support investigation of long-term acceptance, scent habituation, and changing usage patterns over time, while also capturing the social and environmental complexity of everyday work contexts. Future research could further incorporate controlled comparative designs and stronger quantitative evaluation of behavioral, affective, and performance-related outcomes.

\section{Conclusion}
Through an exploratory deployment with 25 knowledge workers, we examined translating physiological stress and cognitive fatigue into ambient olfactory physicalizations for desk-based work. By offering subtle, scene-oriented olfactory metaphors rather than explicit metrics or intrusive alerts, AuraDesk supported low-demand awareness and brief restorative moments without adding notification burden. Our prototype shifts from presenting stress as a readable metric toward interfaces where it can be encountered, interpreted, and gently coped with as part of the everyday sensory atmosphere.

\bibliographystyle{ACM-Reference-Format}
\bibliography{main}
\newpage

\appendix

\section{Post-Study Survey Items}
\label{appendix:survey-items}

The post-study survey was designed to capture key experiential dimensions of interacting with AuraDesk and to make explicit the relationship between the questionnaire items and our research questions. All items were rated on a 7-point Likert scale ranging from 1 (\textit{strongly disagree}) to 7 (\textit{strongly agree}).

To support interpretive clarity, the items were organized into four conceptual dimensions: \textit{Perceptibility and Ambient Quality}, \textit{Interpretability of the Mapping}, \textit{Alignment with Experienced State}, and \textit{Awareness and Regulation}. The first two dimensions primarily addressed RQ1 by examining whether stress-related physiological changes could be translated into ambient olfactory cues that remained both perceptible and interpretable during ongoing work. The latter two dimensions addressed whether participants experienced the cues as personally meaningful and practically useful, and provided a structured basis for reflecting on differences between olfactory feedback and more familiar visual- or notification-based forms of stress communication in the follow-up interviews.

These dimensions were used to organize data collection and descriptive analysis; they were not treated as validated psychometric subscales.

\subsection{Perceptibility and Ambient Quality}
This dimension examined whether scent cues were noticeable, comfortable, appropriately timed, and non-intrusive during ongoing desk work.
\begin{enumerate}
    \item The scent cues were readily noticeable during work.
    \item The scent experience was comfortable overall.
    \item The scent experience was pleasant overall.
    \item The timing of scent releases felt appropriate.
    \item The scent cues were noticeable without being disruptive.
    \item The scent cues felt ambient rather than interruptive.
\end{enumerate}

\subsection{Interpretability of the Mapping}
This dimension examined whether participants could understand the meaning of the scent cues and make sense of the physiological-state-to-scent mapping.
\begin{enumerate}
    \setcounter{enumi}{6}
    \item The meaning of the scent cues was easy to understand.
    \item The physiological-state-to-scent mapping made sense to me.
\end{enumerate}

\subsection{Alignment with Experienced State}
This dimension examined whether participants felt that the scent cues corresponded with their own perceived stress or fatigue.
\begin{enumerate}
    \setcounter{enumi}{8}
    \item The scent cues generally aligned with how I was feeling.
    \item The scent cues appropriately reflected my stress or fatigue state.
\end{enumerate}

\subsection{Awareness and Regulation}
This dimension examined whether the scent cues supported awareness of internal state and were perceived as useful for reflection or self-regulation during work.
\begin{enumerate}
    \setcounter{enumi}{10}
    \item The scent cues helped me become more aware of my stress or fatigue during work.
    \item AuraDesk was useful for supporting awareness and regulation during work.
\end{enumerate}

\section{Technical Implementation of the PicoLM and Actuation Logic}
\subsection{PicoLM Architecture and State Estimation}
To process real-time physiological data on an edge device, AuraDesk utilizes PicoLM, a lightweight sequence-modeling architecture optimized for the NVIDIA Jetson Nano. The model takes a sliding window of physiological features (e.g., RMSSD and SDNN derived from HRV, and baseline-normalized HR) alongside discrete contextual flags (e.g., continuous work duration).
Rather than classifying discrete emotions, PicoLM is trained to project the user's current physiological state onto a continuous 2D Arousal-Valence (A-V) coordinate system. To ensure real-time performance on the edge, the model undergoes post-training quantization (INT8), reducing memory footprint while maintaining sufficient inference accuracy for ambient interaction.

\subsection{The Rule-Based Actuation Matrix}
The continuous A-V coordinates generated by PicoLM are fed into a deterministic rule-based matrix (as shown in Figure 2). This matrix divides the A-V space into distinct interaction zones. For example, a state of high arousal and low valence (indicating elevated stress) triggers the `Forest' profile (e.g., Cedarwood) to support calming. The rule set acts as a critical safety mechanism: it applies temporal smoothing to the AI's output, enforces a minimum 15-minute cooldown between actuations to prevent olfactory fatigue, and translates the magnitude of the A-V deviation into specific hardware commands (PWM duty cycle for intensity, and active duration).

\end{document}